\documentclass[pre,twocolumn]{revtex4}
\usepackage{graphicx}
\usepackage{color}

\begin{document}

\title{Complete wetting of elastically responsive substrates}

\author{N. R. Bernardino}
\affiliation{Centro de F\'{\i}sica Te\'{o}rica e Computacional, Avenida Professor Gama Pinto 2, P-1649-003 Lisboa, Portugal,}
\affiliation{Departamento de F\'{\i}sica, Faculdade de Ci\^{e}ncias da Universidade de Lisboa, P-1749-016 Lisboa, Portugal}

\author{S. Dietrich}
\affiliation{Max-Planck-Institut f\"{u}r Intelligente Systeme, Heisenbergstr. 3, 70569 Stuttgart, Germany,}
\affiliation{Institut f\"{u}r Theoretische und Angewandte Physik, Universit\"{a}t Stuttgart, Pfaffenwaldring 57, 70569 Stuttgart, Germany}

\begin{abstract}

  We analyze theoretically complete wetting of a substrate supporting an array of parallel, vertical plates which can tilt elastically. The adsorbed liquid tilts the plates, inducing clustering, and thus modifies the substrate geometry. In turn, this change in geometry alters the wetting properties of the substrate and, consequently, the adsorption of liquid. This geometry-wetting feedback loop leads to stepped adsorption isotherms with each step corresponding to an abrupt change in the substrate geometry. We discuss how this can be used for constructing substrates with tunable wetting and adsorption properties.

\end{abstract}

\maketitle

\section{Introduction}
\nocite{Rowlinson1982}\nocite{deGennes1985}\nocite{Dietrich1988}
The scientific interest in wetting properties of liquids at solid surfaces dates back to the 18th century. Several reviews account for the progress in understanding such wetting phenomena \cite{Rowlinson1982,Dietrich1988,deGennes1985,Bonn2009}. Recent important developments include the wetting of rough surfaces and superhydrophobicity \cite{Quere2008,Roach2008} and the attempts to build ``lab-on-a-chip'' devices, in which liquids are controlled at the micro- and nanoscale \cite{Squires2005,Rauscher2008}.

A large part of this recent development is concerned with designing surfaces in order to obtain specific, tunable wetting properties, leading to ``smart surfaces". In this spirit non-rigid substrates literally provide additional flexibility in tailoring wetting properties. Recently the usefulness of the interplay of surface elasticity and wetting has been demonstrated in various accounts \cite{Roman2010}. They include capillary-driven self-assembly \cite{Chakrapani2004,Cho2010,Kaur2007,Lim2010,Py2007a,deVolder2010, Kang2011}, enhanced condensation properties \cite{Sokuler2010}, control of surface color \cite{Ch2009a}, measurement of pressure in nano-channels \cite{Tas2010}, and a biomimetic proposal to achieve superhydrophobic surfaces \cite{Otten2004, Mock2005}. This last suggestion has been refuted \cite{Bernardino2010a} but calculations indicate that a related effect might work \cite{Blow2010}. Even at the macroscopic level the interplay of wetting and elasticity gives rise to rich phenomena \cite{Bico2004,Py2007,Boudaoud2007}. 

Here we study an elastic surface structure which changes its geometry upon adsorption. In turn, this change in the topography of the substrate affects the adsorption properties. Our goal is to give a bird's eye view of the influence of elasticity on wetting and adsorption phenomena rather than to calculate specific, detailed results. As such, we choose one of the simplest possible examples of the influence of elasticity: a surface decorated with rigid, vertical, planar plates which can pivot elastically upon their base. This choice simplifies the analysis, while still retaining the important elements of the phenomenology. The adsorption isotherms exhibit jumps at undersaturations at which the surface changes its geometry. The different kinds of possible isotherms depend on elasticity, geometry, and surface tension. We discuss how the use of this kind of surface together with the application of an electric field can lead to ``smart surfaces'' with ``on-the-fly'' tunable wetting properties.

In the next section we introduce the phenomenology of wetting, filling, and capillary condensation and qualitatively describe the wetting of flexible substrates. In Sect. \ref{model} we present the results for our simple macroscopic model. We close with discussions of the results (Sect. \ref{discussion}) and of their implementation in order to obtain surfaces with tunable properties (Sect. \ref{conclusion}).

\section{Phenomenology}
\label{phenomenology}

Before we consider the wetting and adsorption properties of elastic substrates we first briefly review some of the most important results for wetting of planar and patterned substrates (such as wedge filling and capillary condensation), because we rely heavily on these for our calculations.

\nocite{Krim1984}\nocite{Migone1986}\nocite{Bruschi1988}\nocite{Zimmerli1992}\nocite{Hess1997}
In all situations we consider a solid substrate in thermal equilibrium with a bulk gas phase at a certain  temperature $T$ and pressure $P$ (or chemical potential $\mu$), thus considering the grand canonical ensemble. Away from gas-liquid coexistence in the bulk a thin layer of liquid adsorbs on the substrate. The adsorption isotherm is obtained by monitoring the excess adsorption $\Gamma$ upon increasing the pressure towards liquid-vapor coexistence in the bulk at fixed temperature. The excess adsorption can be measured experimentally with high accuracy by using, e.g., microbalance techniques \cite{Bruschi1988,Hess1997,Krim1984,Migone1986,Zimmerli1992}. This is a classical and one of the easiest methods to characterize the wetting properties of a substrate. In the present context we consider the number densities $\rho_l$ and $\rho_v$ of the liquid and vapor phase, respectively, to be spatially homogeneous so that the excess adsorption equals $\rho_l-\rho_v$ times the volume occupied by the liquid phase. Factoring out $\Delta\rho=\rho_l-\rho_v$ and the linear extension of the system in the translationally invariant direction, $\Gamma$ is the area occupied by liquid in the cross section (see Appendix A).

\begin{figure}
  \includegraphics[width=0.5\textwidth]{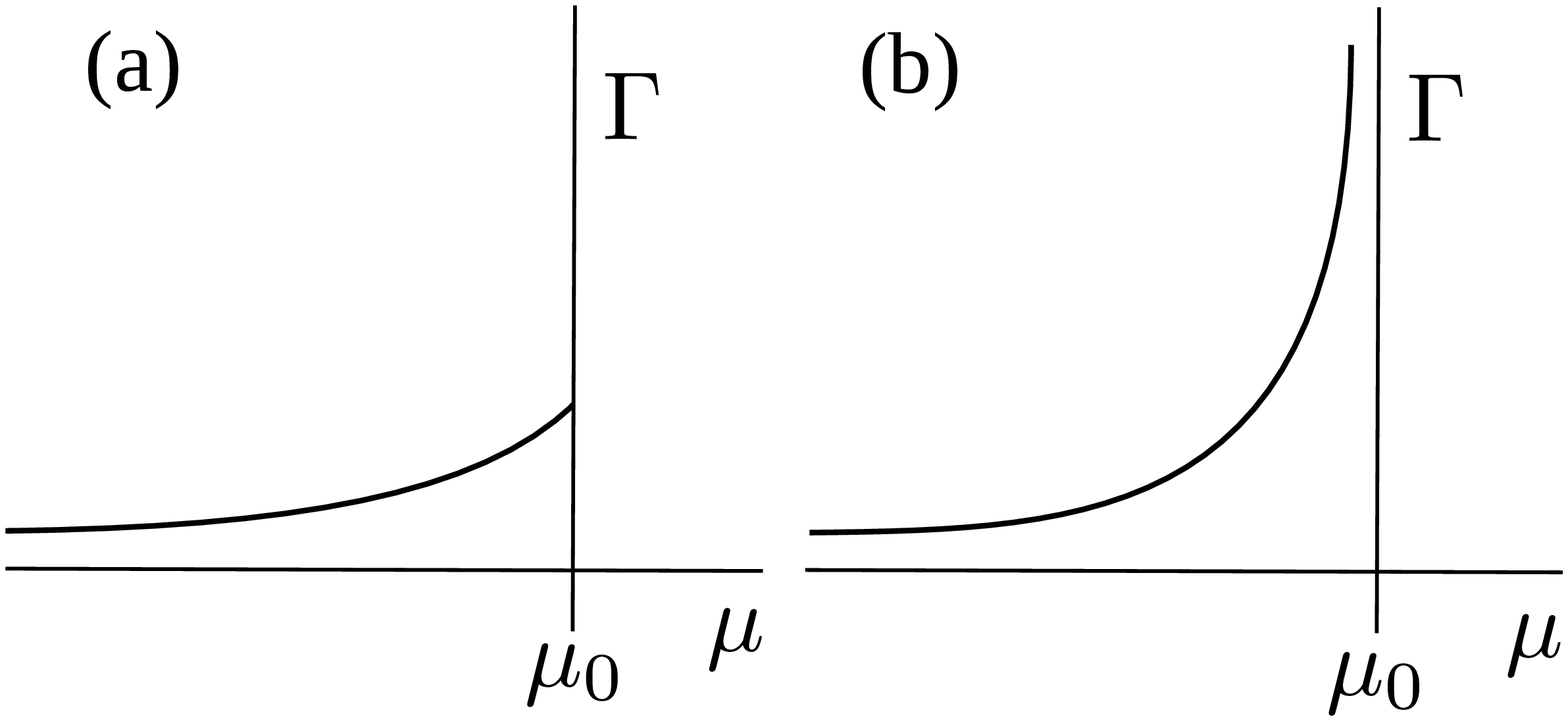}
\caption{\label{isoterms} Characteristic adsorption isotherms of a planar substrate as the chemical potential $\mu$ approaches bulk gas-liquid coexistence $\mu_0(T)$ at constant temperature. $\Gamma$ is the excess amount of liquid adsorbed for partial wetting (a) and complete wetting (b). In (b) the type of divergence of $\Gamma$ reflects the asymptotic decay of the underlying interactions in the system \cite{Dietrich1988}.}
\end{figure}

For a flat substrate there are two typical behaviors of the adsorption isotherms (see Fig.~\ref{isoterms}): either the excess adsorption $\Gamma$ remains finite up to gas-liquid coexistence $\mu_0(T)$ or it diverges as coexistence is approached \cite{Dietrich1988}. The former behavior is called partial wetting, corresponding to a contact angle $\theta>0^\circ$ at gas-liquid coexistence, and the latter is called complete wetting, corresponding to $\theta =0^\circ$. The character of the divergence depends on the microscopic details of the molecular interactions. For non-retarded van der Waals interactions the adsorption diverges as $\Gamma \propto (\Delta\mu)^{-3}$ \cite{Dietrich1988}, where $\Delta\mu \equiv \mu_0(T) - \mu \ge 0$ is the deviation of the chemical potential of the bulk gas phase from its value $\mu_0(T)$ at liquid-vapor coexistence in the bulk.

Due to confinement, fluid will condense in a pore at a lower value of the chemical potential, compared to bulk coexistence. For the simplest possible pore, formed by two planar, parallel, infinite plates a distance $L$ apart, the comparison of the free energies for the empty and the filled pore leads to the Kelvin equation, which provides the value of the chemical potential at which capillary condensation occurs \cite{Dietrich1988}:
\begin{equation}
 \Delta\mu = \Delta\mu^{cc} \equiv \frac{2\sigma \cos\theta}{\Delta\rho L}.\label{kelvin}
\end{equation}
Here $\sigma$ is the surface tension of the liquid-vapor interface.

\begin{figure}
  ~\hfill\includegraphics[height=20ex]{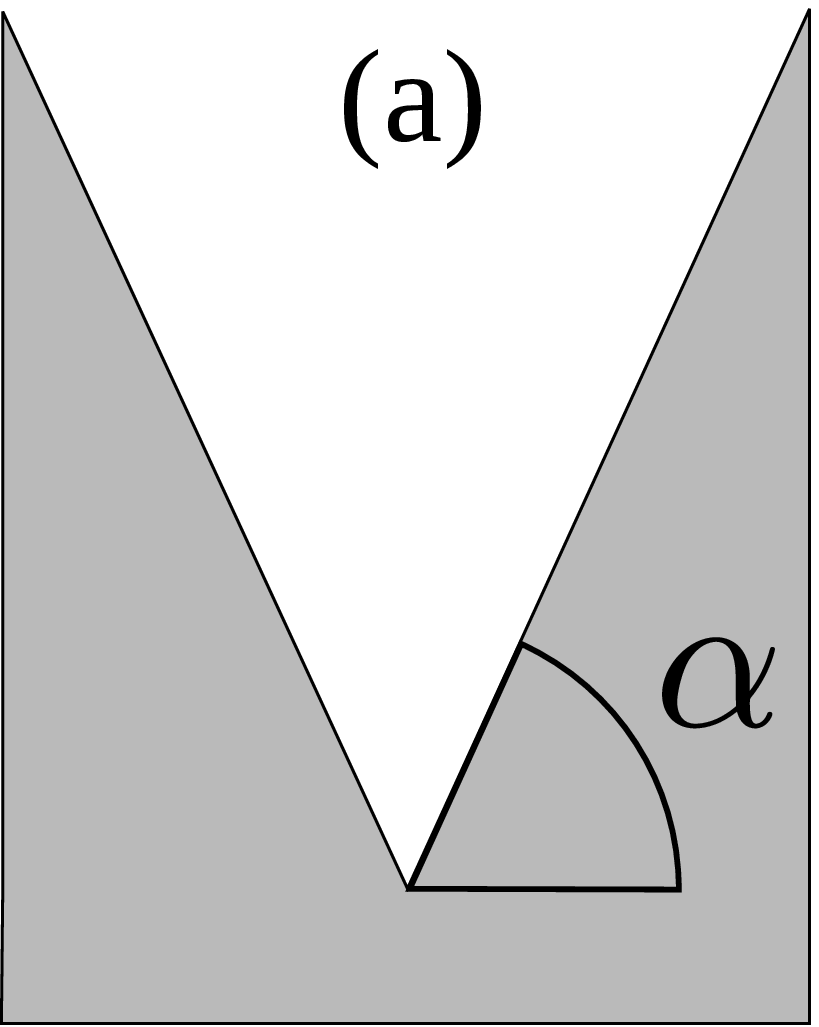}\hfill
  \includegraphics[height=20ex]{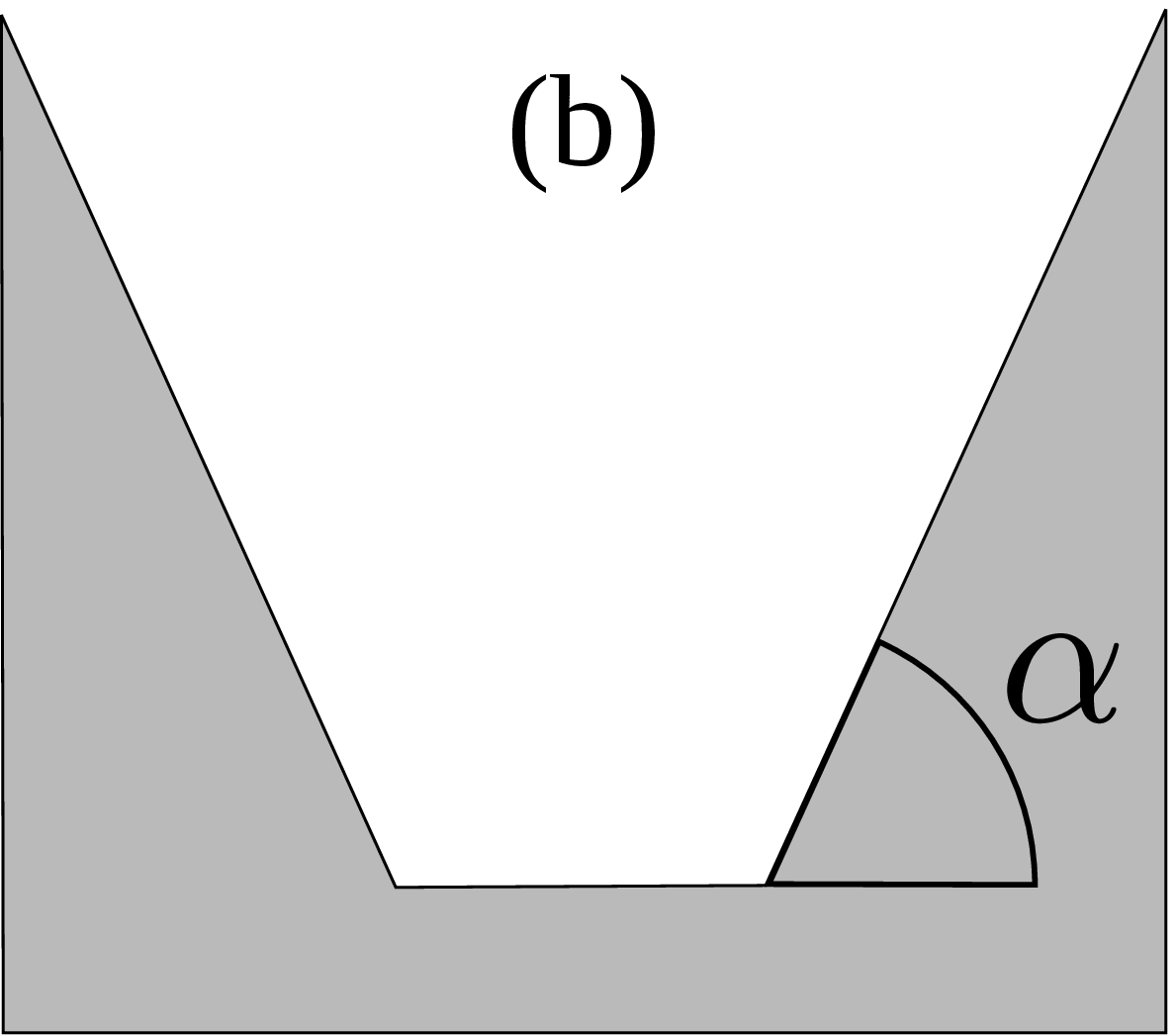} \hfill~
\caption{\label{figwedge}A wedge (a) and a capped wedge (b) with tilt angle $\alpha$.}
\end{figure}

\nocite{Rejmer1999}\nocite{Parry2000a}
Another important element of our later analysis is the adsorption at a wedged substrate with a tilt angle $\alpha$ (see Fig.~\ref{figwedge}(a)). Macroscopic thermodynamic arguments \cite{Rejmer1999,Parry2000a} show that upon raising temperature the filling of a wedge precedes wetting of its walls and occurs when the contact angle at the planar wall equals the tilt angle of the wedge, i.e., $\theta=\alpha$. Insight into the wetting behavior of more complicated substrate geometries can be gained by using a simple mesoscopic model of adsorption \cite{Rasc'on2000}.

\begin{figure*}
  \includegraphics[width=0.95\textwidth]{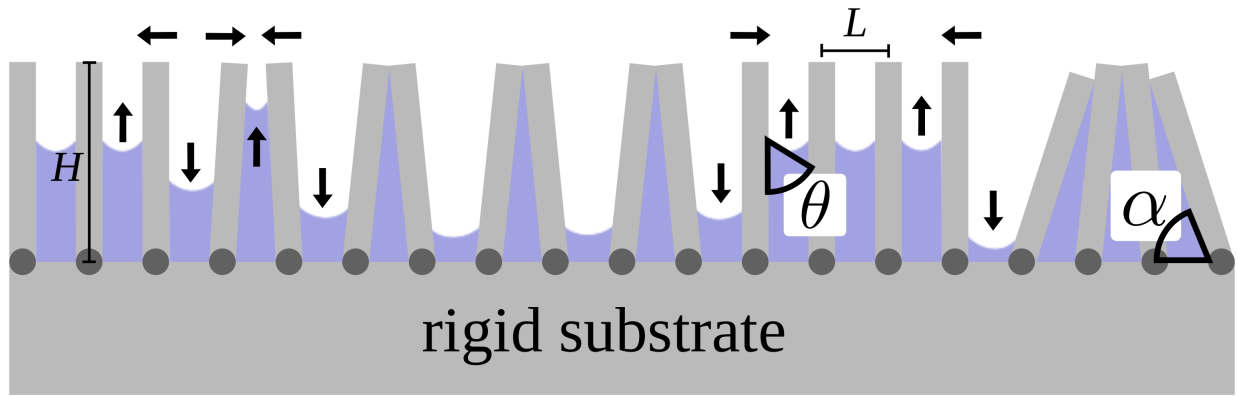}
\caption{\label{substrate}(Color online) Model elastic substrate. The rigid underlying substrate is decorated with rigid, planar, and equally spaced plates. The substrate and the plates are considered to be formed by the same material. The plates can pivot elastically upon their line-like bottom, where they are anchored at the underlying substrate, represented by a dark gray spot. An adsorption-induced collapse of plates triggers a chain of clustering events. The arrows represent the concomitant movement of the liquid-vapor interface and of the plates. Subsequently the clustered plates can collapse into bigger clusters. The sketch exaggerates the thickness of the plates, which are considered to be thin enough, compared to $L$, so that the thickness plays no role in our calculations but thick enough so that microscopic details, such as interactions between portions of liquid adsorbed on different sides of a plate, are not important.}
\end{figure*}

With these prerequisites, we are now in a position to qualitatively describe the adsorption of liquid at an elastic substrate. To be specific, we consider a substrate decorated with a periodic array of planar plates which are vertical if there is no liquid adsorbed (see Fig.~\ref{substrate}). The underlying substrate is rigid as well as the plates as such. However, the plates can tilt elastically around their contact line with the underlying substrate.

\begin{figure}
 \includegraphics[width=0.5\textwidth]{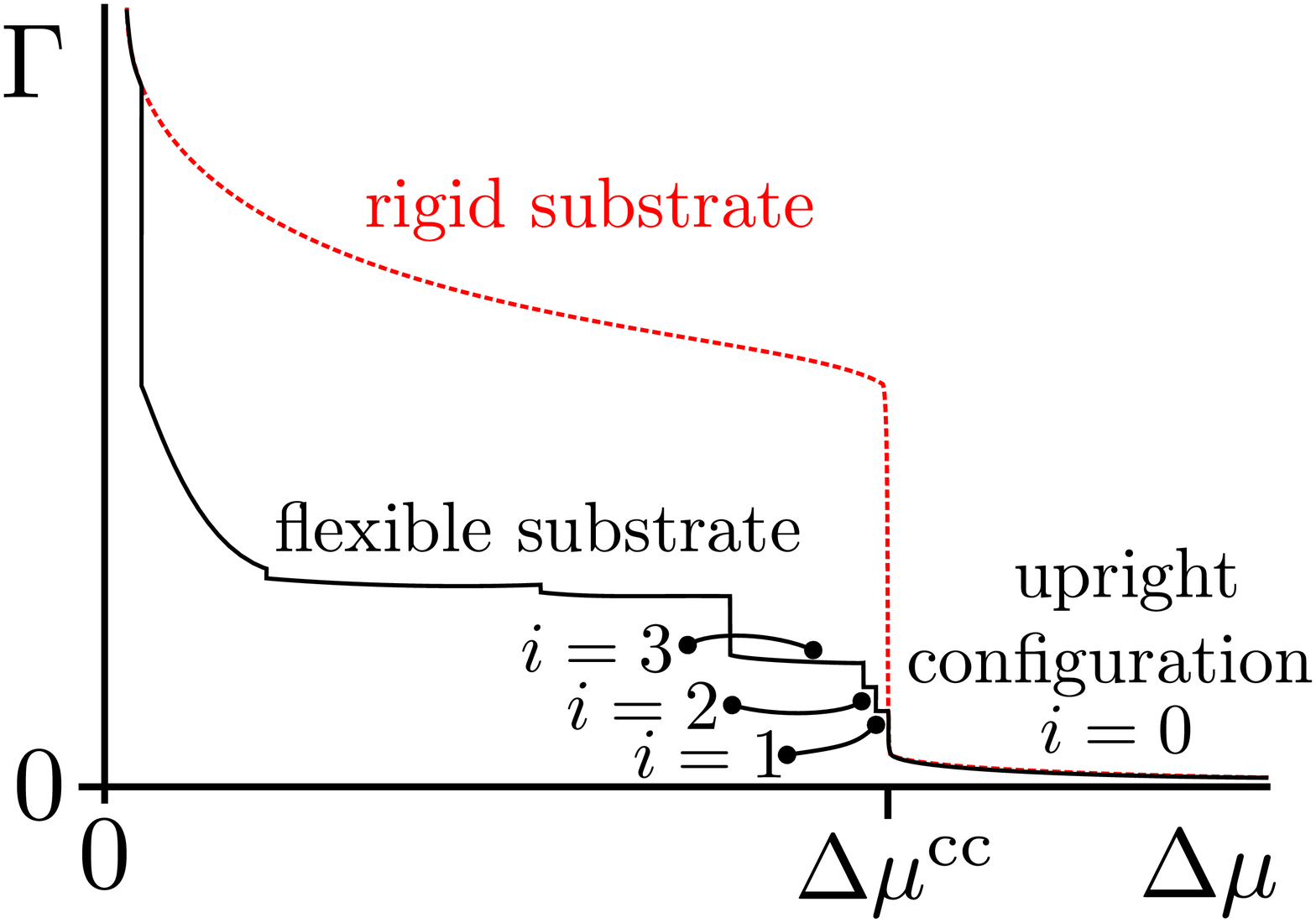} 
\caption{\label{rigid}(Color online) Sketch of the excess adsorption isotherms $\Gamma$ of a rigid, patterned substrate with the plates fixed in upright position (dotted red line) and of a flexible substrate (continuous black line) with the same geometry if there is no adsorption (i.e., $\Delta\mu \gg \Delta\mu^{cc}$). The latter exhibits jumps when the substrate geometry changes, with the order $i$ of the clusters indicated. When the liquid completely covers the plates of the flexible substrate, the plates can return to their original upright position and the adsorption isotherm is the same for $\Delta\mu\to 0$, or they can remain clustered and the isotherm will be different. The plot corresponds to the former case. The substrate material exhibits complete wetting by the liquid (i.e., $\theta=0$). See the main text for a detailed discussion.}
\end{figure}

\nocite{Tasinkevych2006}\nocite{Tasinkevych2007}\nocite{Parry2007}\nocite{Roth2011}
We consider also this substrate to be in thermal equilibrium with a gas phase at a certain temperature $T$ and chemical potential $\mu$ characterizing the grand canonical ensemble. If the substrate is stiff (i.e., if elasticity can be ignored) the excess adsorption isotherm $\Gamma$ typically looks like the dotted line in Fig.~\ref{rigid}. For large undersaturations $\Delta\mu$ only a small amount of liquid adsorbs at the walls, forming a microscopic wetting film. The precise amount of that adsorbed liquid depends on the microscopic details of the system and can be calculated with the same methods used for a planar substrate \cite{Dietrich1988}. At $\Delta\mu=\Delta\mu^{cc}$ the adsorption isotherm exhibits a steep increase at an undersaturation corresponding to the onset of capillary condensation of liquid in the space between the plates. Unlike capillary condensation between parallel walls, this increase is actually smooth, albeit very steep, because the pore is a capped capillary \cite{Parry2007, Roth2011, Tasinkevych2006, Tasinkevych2007}. For even smaller undersaturations the liquid continues to adsorb on the topographically structured substrate with the type of the corresponding divergence being controlled by the range of the molecular interactions involved, similar to a planar substrate \cite{Dietrich1988}. For $\Delta\mu<\Delta\mu^{cc}$ and $\Delta\mu\to 0$ one can identify several filling and wetting regimes \cite{Tasinkevych2006, Tasinkevych2007}.

We now consider elasticity, i.e., the plates can bend due to the capillary forces of the adsorbed liquid.
Starting from large undersaturations, i.e., $\Delta\mu\gg\Delta\mu^{cc}$, upon decreasing $\Delta\mu$ the adsorption isotherms for the elastic substrate are \textit{de facto} the same as the ones for the corresponding rigid structured substrate until capillary condensation occurs. (Concerning flexible slit pores which, however, maintain their geometry, see Refs.~\cite{Ancilotto2011} and~\cite{Schoen2011}.) At the undersaturation $\Delta\mu^{cc}$ corresponding to capillary condensation the liquid starts to fill the space between the plates as before. However, if the elastic constant for tilting a plate around its baseline is not too big, the ensuing capillary forces can cause the plates to cluster together. This can be rationalized by noting that a small deviation of a plate from its vertical orientation causes the pore to be smaller on one side and bigger on the other side; accordingly, these pores form capped wedges (see Fig.~\ref{figwedge}(b)) with tilt angles $\alpha$ smaller or larger than $90^\circ$, respectively. In line with the behavior of wedge filling \cite{Bruschi2002,Parry2000a, Rejmer1999} liquid condenses in the smaller pores while it evaporates from the larger ones, relative to the upright configuration $\alpha=90^\circ$. The condensation and evaporation of liquid in pores of different sizes create an imbalance of forces acting on their common plate due to the surface energies of the interfaces involved. This imbalance triggers a clustering cascade, in which the plates form clusters of two plates each (Fig.~\ref{substrate}), assuming that the plates are not too rigid. This first plate clustering event occurs at the same undersaturation as the aforementioned capillary condensation: $\Delta\mu^{(0)}=\Delta\mu^{cc}$.

At the end of this process the substrate consists of capped wedges, with the plates clustered two-by-two. Between each capped wedge there is a filled wedge upside down (Fig.~\ref{substrate}). Upon decreasing the undersaturation further, adsorption proceeds up to $\Delta\mu=\Delta\mu^{(1)}$ where the filling of these capped wedges causes pairing in clusters of four plates by the same instability mechanism as described above. In line with experimental evidence \cite{Ch2010,Chandra2010,Paulose2010} we assume that, once formed, clusters cannot separate. In this case and at this stage the system consists of capped wedges with a smaller tilt angle $\alpha$, which will in turn collapse into clusters of eight plates (at $\Delta\mu=\Delta\mu^{(2)}$), etc. In our notation, upon decreasing $\Delta\mu$ there is a first-order transition between a cluster of order $i$ (i.e., containing $2^i$ plates) to a cluster of order $i+1$ containing $2^{i+1}$ plates. The upright configuration corresponds to $i=0$.

According to the above description the adsorption properties change upon every clustering event, turning a smooth adsorption curve into a series of adsorption steps which coincide with the abrupt changes of the substrate geometry, as sketched in Fig~\ref{rigid}. These collapses are limited by the length of the plates or by the elastic energy associated with the tilting of the plates, i.e., finally the clusters become too rigid. When no further collapses are possible the adsorption proceeds by filling of the capped wedges (see the smooth increase of $\Gamma$ in Fig.~\ref{rigid} below the last step) until the liquid completely fills the space between the plates. 
Therefore the value of $\Gamma$ at the bottom end of the last step (corresponding to a filled capped wedge configuration) equals the one at the upper end of the steeply increasing dotted line corresponding to a filled upright configuration.
Further adsorption causes the liquid to submerge the plates and the surface tension of the liquid-vapor interface no longer acts on the plates. The plates can remain stuck to each other if van der Waals or other forces keep them together. Alternatively, elasticity can be strong enough to restore the plates to their upright configuration. In the former case further adsorption is similar to adsorption on a substrate formed by capped wedges, whereas in the latter case further adsorption is the same as on the rigid substrate, with the plates in the upright configuration; this is the case drawn in Fig.~\ref{rigid}. In either case, for the slope of this part of the adsorption isotherm elasticity does no longer play a role.

If the substrate is not perfectly periodic but there is some randomness in the spacing of the plates the adsorption will proceed in a less orderly fashion \cite{Kang2011}. For example, there can be clusters of three plates, or clusters of different sizes within the overall configuration. This will result in more steps in the adsorption isotherm.

In the next section we present results for a specific model system, substantiating the previous qualitative reasoning.

\section{Model Calculations}
\label{model}

In order to underpin the previous phenomenological description, in this section we present  calculations for a simple model which still captures the most important features but avoids the complexity of more realistic models.

As stated before our choice of the substrate is idealized in that it consists of stiff planar plates which tilt around their line of contact with the underlying substrate. A more realistic system would consist of elastic plates which can adopt also nonplanar shapes. This would lead to much more difficult calculations without adding essential new features. Moreover, the system we analyze is translationally invariant along one spatial direction, thus being effectively two-dimensional. This invariance is broken by studying systems with arrays of elastic posts. Such an analysis would be rather challenging because it requires one to solve the Laplace equation within a domain with a complicated shape in order to obtain the equilibrium configuration of the liquid-vapor interface. In the following we adopt this translational invariance and all extensive quantities are taken to be divided by the length of extension in this direction. For example the expression for the energy of the system (see, c.f., Eq.~(\ref{Etotal})) denotes the energy per length.

For our simplified model we introduce the additional approximation that the top of the plates is always at a vertical distance $H$ from the substrate. This approximation is justified if the height $H$ of the plates is much larger than the distance $L$ between vertical plates and if the clusters do not involve too many plates (see Fig.~\ref{substrate}).

In order to calculate the adsorption we resort to a macroscopic model, in which the liquid-vapor interface is a section of a cylinder with its radius of curvature $R$ given by Laplace's equation:
\begin{equation}
 R = \frac{\sigma}{\Delta\rho\Delta\mu}.
\label{laplace}
\end{equation}
(A more sophisticated model would include a thin layer of liquid adsorbed on the walls \cite{Rasc'on2000}. In a first approximation we neglect this.) Our macroscopic (thermodynamic) approach implies that, for the non-tilted plates, only at capillary condensation the liquid meniscus forms a part of a circle, the radius of which is given by Eq.~(\ref{laplace}) and which meets the vertical wall with the contact angle $\theta$. Actually, this is precisely what signals the occurrence of capillary condensation.

The free energy of the system for a given configuration consists of a volume term (capturing that the system is off liquid-vapor coexistence), surface terms, and the elastic energy of tilting the plates. In principle this free energy has to be minimized in order to obtain the equilibrium configuration of the system and, from this, the adsorption isotherms. However, the implementation of this procedure is rather complicated. Moreover, for these types of systems it is not obvious that they are always able to reach the free energy minimum \cite{Bernardino2010,Ch2010}. Therefore we use instead a more ad-hoc approach, in analogy with  Ref. \cite{Bernardino2010a}.

Our aim is to determine the values of the chemical potential for which the clustering of the plates occurs. In between these values for the clustering events we treat the substrate as to be rigid and describe the adsorption without considering elasticity. To be specific, we calculate the values  $\Delta\mu^{(0)},\Delta\mu^{(1)},\Delta\mu^{(2)},...$  of the chemical potential for which the first, second, third, and so on clustering event occurs. The adsorption between these values of the chemical potential is taken to be the same as the adsorption on a rigid substrate with the plates fixed to be clustered correspondingly, i.e., two-by-two, four-by-four, etc.

The first step is to determine if the plates cluster at all. According to the Kelvin equation \cite{Dietrich1988} capillary condensation between two parallel plates occurs at $\Delta\mu = \Delta\mu^{cc}$ (see Eq.~\ref{kelvin}). This equation is valid for condensation between two infinite parallel plates. For a finite pore, i.e., for a capped capillary, the condensation transition still occurs for the same undersaturation $\Delta\mu^{cc}$ but the transition changes from first-order to continuous (albeit with a very steep increase of the adsorption) \cite{Parry2007}. Once again, the accurate description of this transition requires a more detailed model but this is of no relevance for the present analysis.
In the following we consider thin plates, the thickness of which is much smaller than $L$ which allows us to neglect it. However, capillary condensation does not occur in all pores simultaneously because as soon as one pore fills, the surface tension pulls its confining plates together, causing the neighboring pores to open and thus preventing condensation in these. Therefore, in order to determine if the plates cluster, one has to balance the elastic, the volume, and the surface tension energies between two individual plates filled by the liquid up to the top:
\begin{equation}
\label{instability}
 E = 2k(\pi/2-\alpha)^2 + \Delta\rho\Delta\mu H (L-x) + \sigma(L-2x).
\end{equation}
Here $k$ is the elastic constant, $\alpha$ is the angle the plate forms with the horizontal, and $x$ is the deviation of the top of the plate from the reference configuration $\alpha=\pi/2$; $x>0$ $(x<0)$ refers to closing (opening)  the gap. Equation~(\ref{instability}) assumes a planar instead of a bent liquid-vapor interface at the top of the pore, which represents a minimal requirement for the collapse of the plates, i.e., if the pores are not completely filled up to the rim the area of the liquid-vapor interface is larger. Therefore the configuration of a flat interface at the top is the one with the largest possible, bending induced, free energy reduction associated with the liquid-vapor interface. This implies that if the surface tension cannot cause the bending of the plates even when the interface reaches up to the top everywhere, it never will lead to bending.

Defining $\delta\alpha = \pi/2-\alpha$ and using $\delta\alpha\approx \tan(\delta\alpha) = x/H$ and solving for $dE/dx=0$ we obtain $x^* = \frac{\sigma H^2}{2k} + \frac{\Delta\rho\Delta\mu H^3}{4k}$.
If $x^*<L/2$ the plates bend due to the capillary forces but do not touch each other and, as the chemical potential is increased further (i.e., $\Delta\mu$ is decreased), the neighboring pores with a wide opening fill without any further clustering. On the other hand, if the formal solution $x^*$ of the equation $dE/dx=0$ fulfills $x^*>L/2$, it is outside of the physically allowed domain for $x^*$; in fact this means that the plates would have to ``cross'' each other. Thus the condition $x^*>L/2$ tells that the energy is minimized if the plates bend until they touch each other, at which point a hard-body repulsion prevents the ``crossing'' of plates.
The above dependence of $x^*$ on $\Delta\mu$ deserves a comment, as it might appear to be surprising that a larger $\Delta\mu$ leads to a larger $x^*$. One must keep in mind that this relationship is valid under the assumption that the liquid-vapor interface is stuck at the top of the plates. The increase of $\Delta\mu$, while forcing the interface to stay at the top of the plates, causes the plates to bend as much as possible in order to reduce the volume of liquid as it is required for large undersaturations. However, this is not important for our analysis which is focused on the behavior of $x^*$ only near capillary condensation.
Therefore the necessary condition for two individual plates to cluster is
\begin{equation}
 L < \frac{\sigma H^2}{k} + \frac{\Delta\rho\Delta H^3}{2k}.
\label{stability}
\end{equation}

After this first clustering, the system is formed by capped wedges. If we assume that in this first step ($i=1$) the plates cluster pairwise, then by four $(i=2)$, followed by eight $(i=3)$ plates, and so on, the outermost tilt angle after the clustering of the $i^{\text{th}}$ order is
\begin{equation}
 \tan \alpha_i = \frac{2H}{(2^i-1)L}.
\label{angle}
\end{equation}
Note that $\alpha_{i=0}=\frac{\pi}{2}$.


In the following we first consider a system of \emph{fixed} clusters of order $i$. In order to obtain the corresponding excess adsorption $\Gamma_i$ we simply add the volumes $\Gamma_{\text{c}}^{(i)}$ and $\Gamma_{\text{w}}^{(i)}$ of liquid in a cluster of order $i$ and  in the remaining wedges, respectively, multiplied by the total number $N_{\text{c}}^{(i)}$ of clusters  of order $i$ in the system with $N$ plates where $N_{\text{c}}^{(i)}=N/2^i$ after the $i^{\text{th}}$ collapse. We have
\begin{equation}
\label{gammac}
 \Gamma_{\text{c}}^{(i)} = \frac{H^2}{\tan\alpha_i},
\end{equation}
which, within the present simple model, is independent of $\Delta\mu$. Thus we consider the situation that fluid trapped within a pore does not evaporate if $\Delta\mu$ is increased. (To a certain extent this lack of evaporation upon increasing $\Delta\mu$ mimics hysteresis.) A more refined model would take this evaporation into account, but this has no consequence for our main results. For $\Gamma_{\text{w}}^{(i)}$ we have (see Eq.~(\ref{laplace}) and Eq.~(\ref{wedge}) in Appendix~\ref{append})
\begin{equation}
 \Gamma_{\text{w}}^{(i)} = \frac{\sigma^2}{(\Delta\rho\Delta\mu)^2}\left[ \frac{\sin(\alpha_i-\theta)\cos\theta}{\cos\alpha_i} - (\alpha_i-\theta) \right] -\frac{L^2}{4}\tan\alpha_i.
\label{gammaw}
\end{equation}
The expression for $\Gamma_{\text{w}}^{(i)}$ vanishes for $\Delta\mu \nearrow \Delta\mu^{(i)}_-\equiv \sqrt{\frac{4\sigma^2}{(\Delta\rho L)^2\tan\alpha_i}\left[ \frac{\sin(\alpha_i-\theta)\cos\theta}{\cos\alpha_i} - (\alpha_i-\theta) \right]}$. This signals that for $\Delta\mu > \Delta\mu_-^{(i)}$ there is only a microscopically thin adsorbed film which is not captured by the present macroscopic description. Accordingly the expression in Eq.~(\ref{gammaw}) is augmented by $\Gamma_{\text{w}}^{(i)}(\Delta\mu > \Delta\mu_-^{(i)})=0$.

A more convenient quantity is the adsorption per unit area, i.e., per length if one divides by the extension of the system in the translationally invariant direction:
\begin{equation}
 \frac{\Gamma_i}{NL} = (\Gamma_{\text{c}}^{(i)}+\Gamma_{\text{w}}^{(i)})\frac{N_{\text{c}}^{(i)}}{NL} = \frac{\Gamma_{\text{c}}^{(i)}+\Gamma_{\text{w}}^{(i)}}{2^i L}.
\end{equation}

Now we consider flexible plates. The clustering events occur if a configuration becomes linearly unstable. In order to calculate the value $\Delta\mu^{(i)}$ of the chemical potential at which a given configuration $(i)$ leads to clustering of order $(i+1)$ we check the stability of the system against a small perturbation which tilts the clusters an angle $\delta\alpha$ towards each other.
We consider the situation that the clusters tilt in pairs, so that
the total energy $E_{\text{tot}}^{(i)}(\delta\alpha)=\frac{N_{\text{c}}^{(i)}}{2}\left[ E_i(\alpha_i+\delta\alpha)+E_i(\alpha_i-\delta\alpha) \right]$ is given by (see Eq.~(\ref{energycluster}) in Appendix~\ref{append})
\begin{widetext}
\begin{eqnarray}
 \frac{2^{i+1} E^{(i)}_{\text{tot}}(\delta\alpha)}{N} & =  & \frac{\sigma^2}{\Delta \rho \Delta \mu}
    \Bigg[ - \frac{\cos\theta}{\cos(\alpha_i+\delta\alpha)}\sin(\alpha_i+\delta\alpha-\theta) - \frac{\cos\theta}{\cos(\alpha_i-\delta\alpha)}\sin(\alpha_i-\delta\alpha-\theta)\nonumber\\
    & & - \frac{L^2}{4}\tan(\alpha_i+\delta\alpha)\frac{(\Delta\rho\Delta\mu)^2}{\sigma^2} - \frac{L^2}{4}\tan(\alpha_i-\delta\alpha)\frac{(\Delta\rho\Delta\mu)^2}{\sigma^2}\label{Etotal}\\
    & & + L\frac{\cos\theta}{\cos(\alpha_i+\delta\alpha)}\frac{\Delta\rho\Delta\mu}{\sigma} + L\frac{\cos\theta}{\cos(\alpha_i-\delta\alpha)}\frac{\Delta\rho\Delta\mu}{\sigma}
     \Bigg] +  2^{i+1}k(\delta\alpha)^2 + C^{(i)}(\alpha_i),\nonumber
\label{Energy}
\end{eqnarray}
\end{widetext}
where the last contribution contains terms which do not depend on $\delta\alpha$ and thus can be dropped upon varying $\delta\alpha$.

For each $i$ one has to check if the configuration $\delta\alpha=0$ is linearly stable. By symmetry $\delta\alpha=0$ is always an extremum of $E_{\text{tot}}^{(i)}$. If it becomes linearly unstable the plates form clusters of order $i+1$. This loss of stability can occur smoothly, with the plates tilting gradually towards each other, or abruptly, with the plates clustering instantaneously. Numerically we have always observed an abrupt clustering of the plates.

It is too difficult to analytically determine the point at which the configuration with $\delta\alpha=0$ is no longer linearly stable, i.e., $\left. \frac{d^2 E_{\text{tot}}^{(i)}}{d\delta\alpha^2}\right|_{\delta\alpha=0}\!\!=0$. However, for given values of the parameters we can proceed numerically to determine both the values $\Delta\mu^{(i)}$ of the chemical potential for the clustering events and the isotherms.

Clustering and adsorption can proceed as long as the wedges between clusters are not filled by liquid. For a fixed cluster, this ceases if $h-h'=H$ (see, c. f., Fig.~\ref{diagram}), i.e., at
\begin{equation}
  \label{end}
 \Delta\mu = \Delta\mu^*_i \equiv \frac{\sigma\sin(\alpha_i-\theta)}{\Delta\rho(H/\tan\alpha_i + L/2)}
\end{equation}
with $\alpha_i$ given by Eq.~(\ref{angle}). For $\Delta\mu<\Delta\mu^*_i$ the liquid completely covers the plates. In this case the surface tension does no longer promote the clustering of the plates because the liquid-vapor interface is above the plates. Accordingly the elastic forces will break the clusters apart, unless other forces, such as van der Waals forces, keep them together.

In is also worthwhile to point out that the above description automatically takes into account the possibility that, beyond a certain number of clustering events, the clusters are too rigid to be bent by the surface tension. In such a case the geometric configuration never becomes unstable and the liquid adsorbs until it completely covers the plates, as discussed at the end of Section~\ref{phenomenology}.

\section{Discussion}
\label{discussion}
\begin{figure*}
 \includegraphics[width=0.95\textwidth]{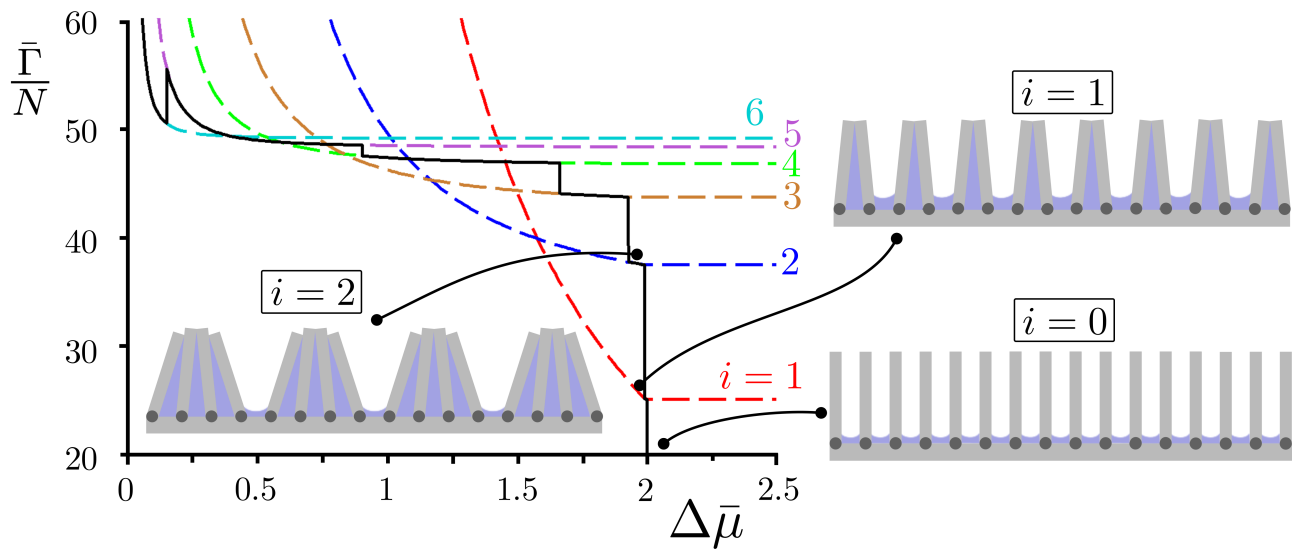} 
\caption{\label{adsorption} (Color online) Adsorption isotherm of an \emph{elastic}, patterned substrate (black line) and isotherms for substrates composed of \emph{rigid}, clustered plates (dashed lines) with clusters of the order $i=1, ..., 5$ from right to left at the top, and from bottom up at the right. The first three configurations of the substrate are sketched, along with pointers indicating the $\Delta\bar\mu$ intervals within which they occur. The clusters of order $i$ containing $2^i$ plates are stable for $\Delta\bar\mu^{(i)}<\Delta\bar\mu<\Delta\bar\mu^{(i-1)}$ (with $\Delta\bar\mu^{(-1)}\equiv\infty$ for $i=0$). The collapse of the clusters of order $i$ takes place at $\Delta\bar\mu^{(0)}=2$, $\Delta\bar\mu^{(1)}=1.990$, $\Delta\bar\mu^{(2)}=1.928$, $\Delta\bar\mu^{(3)}=1.663$, $\Delta\bar\mu^{(4)}=0.901$, and $\Delta\bar\mu^{(5)}=0.148$. The first two collapses occur very close to each other and consequently they are barely discernible in the isotherm. The systems are characterized by the dimensionless parameters $\bar L=1$, $\bar H=100$, $\bar k=50$, and $\theta=0$. See the main text for a detailed discussion.}
\end{figure*}

In order to translate the previous calculations into specific results for the adsorption isotherms it is convenient to introduce dimensionless quantities, which we denote by an overbar such as $\bar E$. To this end we measure the lengths in units of $L$, the energy per length in units of $\sigma L$ (the translational invariance of the system has been already taken into account) and the chemical potential in units of $\frac{\sigma}{\Delta\rho L}$. The elastic constant is also measured in units of $\sigma L$ and the adsorption is measured in units of $L^2$. With this choice of units we are left with the parameters $\bar H \equiv H/L$, which defines the geometry, $\theta$, and $\bar k \equiv\frac{k}{\sigma L}$ which is a ratio of elastic and surface energies. It is not easy to estimate $\bar k$ for actual materials as we are assuming the plates to tilt around their base. In any case a large value of $\bar k$ signals a rigid substrate, for which the surface tension is not capable of bending the plates. On the other extreme, a soft substrate is described by a small value of $\bar k$ such that the surface tension forces cause extensive clustering. The spectrum of actual materials spans the entire range of values, as can be inferred from experimental results, according to which clustering can either occur or not \cite{Chakrapani2004,Cho2010,Kaur2007,Lim2010,Py2007a,deVolder2010, Roman2010, Ch2009a, Cambau2011}. It is to be expected that the interplay of elasticity and surface tension will be most pronounced for intermediate values of $\bar k$. Accordingly, in Fig.~\ref{adsorption} we present our results for the adsorption isotherms (in dimensionless units as described above) for $\bar k=50$, $\theta=0$, and $\bar H=100$. For these parameters the necessary condition for clustering of two individual plates to occur, $\displaystyle 1<\frac{\bar H^2}{\bar k}+\Delta\bar\mu\frac{\bar H^3}{2\bar k}$ (see Eq.~\ref{stability}), is fulfilled for all $\Delta\mu$, because  $1<\frac{(100)^2}{50}+\Delta\bar\mu\frac{(100)^3}{100}$.

Fig.~\ref{adsorption} shows the adsorption isotherms for the first few \emph{fixed} cluster configurations ($i=1,...,6$) with $\bar\Gamma_i/N = 2^{-i}(\bar\Gamma^{(i)}_{\text{c}} + \bar\Gamma^{(i)}_{\text{w}})$
where $\bar\Gamma^{(i)}_{\text{c}}$ and $\bar\Gamma^{(i)}_ {\text{w}}$ are given by Eqs.~(\ref{angle})-(\ref{gammaw}). In line with the discussion after Eq.~(\ref{gammaw}) these isotherms are flat for $\Delta\bar\mu \ge \Delta\bar\mu_{-}^{(i)}$ with plateau values $\frac{2^i-1}{2^{i+1}}\bar H\to\bar H/2$ as $i\to\infty$.
For decreasing undersaturation the isotherms increase up to $\Delta\bar\mu^*_i (\theta=0) = 2^{1-i}\xi_i(1+\xi_i^2)^{1/2}$, $\xi_i=\frac{2\bar H}{2^i-1}$, (see Eqs.~(\ref{angle}) and (\ref{end})) where the wedges become completely filled up and the cluster configurations dissolve; $\Delta\mu_{i\to\infty}^*(\theta=0)=4^{-i}\bar H \to 0$. In Fig.~\ref{adsorption} this latter feature occurs outside the field of view (e.g., $\Delta\bar\mu^*_1 \approx 1$ and $\bar\Gamma_1(\Delta\bar\mu_1^*)/N>60$). When the clusters dissolve the adsorption jumps to the same value as for the rigid substrate, as sketched in Fig.~\ref{rigid}, because elasticity no longer plays a role.

If the cluster configurations are not fixed but flexible, at certain undersaturations $\Delta\bar\mu^{(i)}$ (determined by the linear stability analysis criterion $\frac{d^2 E^{(i)}_{\text{tot}}}{d \delta \alpha^2}=0$, see Eq.~(\ref{Energy}) above) the substrate configuration undergoes a first-order transition between clusters of order $i$ (i.e., containing $2^i$ plates) and clusters of order $i+1$ upon decreasing $\Delta\bar\mu$, with corresponding jumps in the adsorption (see the black line in Fig.~\ref{adsorption}). In between two jumps the adsorption isotherm of the flexible substrate follows the isotherm of the corresponding fixed cluster. This leads to the step-like character of the black line in Fig. \ref{adsorption}

Due to the choice $\theta=0$ capillary condensation occurs at ${\Delta\bar\mu^{cc}} = 2$ (Eq.~(\ref{isoterms})). Thus for the present system clustering sets in for this value of the chemical potential because, as stated above, Eq.~(\ref{stability}) is fulfilled. Therefore one has $\Delta\mu^{(0)}=\Delta\mu^{cc}$.
The significance of elasticity is highlighted by a comparison of the black line in Fig.~\ref{adsorption} with the adsorption isotherm expected for the corresponding rigid substrate featuring the same plates but fixed in upright position as sketched in Fig.~\ref{rigid}. At capillary condensation ($\Delta\bar\mu = \Delta\bar\mu^{cc} = 2$) the adsorption for the rigid substrate jumps suddenly from a small value to $\bar\Gamma/N=100$ which is outside the field of view of Fig.~\ref{adsorption}, corresponding to the complete filling of the space between the plates. Our macroscopic model does not take into account the microscopic details of the fluid-solid and fluid-fluid interactions and therefore it cannot describe the precise shape of this isotherm, only the position of the jump at capillary condensation, predicting for the present choice of parameters constant values of $0$ and $100$ for the adsorption below and above the capillary condensation. A more detailed model (see, e.g., Refs. \cite{Rasc'on2000, Parry2007, Tasinkevych2006, Tasinkevych2007}) is required in order to be able to determine the full shape of the isotherm.

%
%
%
%

The isotherms for elastic substrates differ qualitatively from the isotherms of substrates composed of rigid clusters of fixed order $i$ with a tilt angle given by Eq.~(\ref{angle}) (see also Fig.~\ref{rigid}). As expected, the adsorption isotherm of the elastic surface exhibits abrupt jumps at each clustering event (i.e., at $\Delta\mu=\Delta\mu^{(i)}$), reflecting the change of the substrate geometry. The first two clustering events occur at undersaturations $\Delta\mu$ very close to each other, because the large ratio $H/L=100$ implies that the angles $\alpha_i$ of the wedges formed in the first clustering events are very close to $90^\circ$ and thus in Fig.~\ref{adsorption} the differences between these values of $\Delta\mu^{(i)}$ are barely visible. Another notable feature is that for decreasing $\Delta\mu$ the adsorption is not necessarily monotonically increasing. This counter-intuitive behavior is due to liquid evaporating (as indicated by the downward arrows in Fig.~\ref{substrate}) from the capped wedges, which form when a cluster collapses upon decreasing $\Delta\mu$. An experiment in which the chemical potential is increased slowly, in order to facilitate thermal equilibrium and evaporation of liquid into the vapor phase, is expected to pick up this peculiarity.

It is also informative to monitor the filling height $h-h^{'}-R(1-\cos(\alpha-\theta))$ of the liquid in the center of the capped wedges (see Fig.~\ref{diagram} in Appendix A) as function of the order $i$ of clustering (i.e., at $\Delta\mu^{(i)}$). For $i=1$ to $5$ the heights in units of $L$ are $0.001, 0.73, 2.31, 7.1$, and  $34.2$ compared with the maximal value $H/L=100$. The first few values are rather small, reflecting the fact that $\bar k=50$ describes rather flexible substrates which cluster easily. For these low orders, the plates cluster almost as soon as an interface is formed.

\begin{figure}
 \includegraphics[width=0.3\columnwidth]{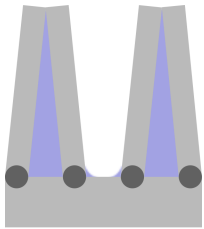} 
\caption{\label{corner}(Color online) Liquid-vapor interface which does not span the space between two clusters but which connects both plates with the substrate surface. Our model does not take into account the adsorption in the corners, but this is important for extremely flexible substrates.}
\end{figure}

As a final comment we point out that our choice of $\bar k=50$ was not accidental. For a smaller value of $\bar k$ the substrate is so flexible that clustering occurs even before the interface spans the space between two plates. The formation of the interface associated with the liquid adsorbed in the corner formed at the point of contact of the plates with the substrate is sufficient to cause clustering. A signature of this feature is that the filling height $h-h^{'}-R(1-\cos(\alpha-\theta))$ discussed in the previous paragraph becomes negative if one applies the approach described before. This shortcoming for small $\bar k$ could be fixed easily by taking into account  the adsorption in the corners formed between the plates and the underlying substrate (see Fig.~\ref{corner}). However, this extra step would not add any qualitatively new feature and would complicate our simple analysis unnecessarily.

\section{Conclusions}
\label{conclusion}

Our analysis shows how a responsive substrate and the coupling between wetting and substrate geometry can lead to new phenomena. For the specific system considered here the adsorption isotherms exhibit  steps at undersaturations for which the substrate changes geometry due to the adsorption of liquid.

For these phenomena adsorption isotherms are particularly revealing in that they are directly accessible to experiments, as already proven for geometrically patterned substrates \cite{Hofmann2010}, and they provide an expressive characterization of the wetting properties of substrates.
Our choice of the substrate might not be the easiest to be prepared experimentally but in view of great advances in the fabrication of patterned substrates \cite{delCampo2008}, it might be possible to fabricate such a substrate or a similar one. A substrate with floppy plates could be created more easily, but the corresponding theoretical analysis will be more difficult, because in this case the bent plates will not form straight wedges but rather more complicated shapes.

We have used a simple macroscopic model in order to calculate the adsorption. It is likely that for a direct quantitative comparison with possible experimental adsorption isotherms our simple model is insufficient. A more accurate model should incorporate more realistic details such as van der Waals interactions \cite{Hofmann2010} and more realistic elastic features of the patterned substrates. Such refined descriptions will be most successful if they are tailored to specific substrates used in actual  experiments.

We end with mentioning a prospective application of our results. So far we have considered ``passive'' substrates. The geometry changes with adsorption and in turn this change is determined by the amount of liquid adsorbed and by the characteristics of the substrate fixed via its fabrication. One can instead imagine a more ``active'' substrate by using, e.g., electric fields in order to control how the substrate structures collapse. Indeed electric \cite{Hill2010} and magnetic fields \cite{Zhou2011} have already been used for control of the collapse of elastic substrates. In this setup one could change externally the effective stiffness of the substrate, and thus control its wetting properties. Such controllable surfaces might find applications in, e.g., ``lab-on-a-chip'' devices.

\begin{acknowledgments}

NRB acknowledges support from the Portuguese Foundation for Science and Technology (SFRH/BPD/63183/2009 and PEst-OE/FIS/UI0618/2011) and thanks Mykola Tasinkevych, Roland Roth, and Jan Guzowski for many fruitful discussions. We also thank Paulo Teixeira for a critical reading of the manuscript.

\end{acknowledgments}

\appendix

\section{Expression for the free energy}
\label{append}

\begin{figure}
  \includegraphics[width=0.5\columnwidth]{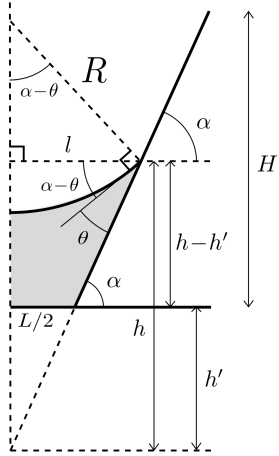}
\caption{\label{diagram}Geometry of the problem depicting the relevant lengths and angles. The gray area denotes the liquid. The actual surfaces are indicated by thick black lines. One has $l=R\sin(\alpha-\theta)$, $h'=\frac{L\tan\alpha}{2}$, and $h=l\tan\alpha$. Note that for a given geometry (i.e., $\alpha$ and $L$) and given materials ($\theta$) the filling height $h-h' -R(1-\cos(\alpha-\theta))$ of the liquid in the center-line of the pore is determined by the radius $R$ of curvature, which in turn is given by the undersaturation $\Delta\mu$ (see Eq.~(\ref{laplace})).}
\end{figure}

Within our model the calculation of the free energy for a given geometric configuration and chemical potential reduces to geometrical considerations. The free energy is the sum of volume, surface, and elastic contributions:
\begin{widetext}
\begin{eqnarray}
 E & = & \Delta \rho \Delta \mu \times (\text{volume of liquid}) + \sigma\times(\text{liquid-vapor surface area})\nonumber\\
 & + & \sigma_{\text{ls}}\times(\text{liquid-solid surface area}) + \sigma_{\text{vs}}\times(\text{vapor-solid surface area})\nonumber\\
& + & \text{elastic energy}.\label{energy}
\end{eqnarray}
\end{widetext}
Equation (\ref{energy}) assumes that the bottom of the cavity and the plates consist of the same solid material so that both give rise to the same surface tension $\sigma_{\text{ls}}$. Fig.~\ref{diagram} illustrates the geometry of the problem and defines all relevant lengths and angles. In the following we omit the linear extension of the system in the translationally invariant direction.

Within a harmonic approximation the elastic energy of a single bent plate forming an angle $\alpha$ with the horizontal substrate is
\begin{equation}
 E_{\text{el}} = k (\pi/2 - \alpha)^2,
\end{equation}
so that the elastic energy of a cluster of the order $i$ is
\begin{equation}
 E_{\text{el}}^{(i)} = 2 k \sum_ {j=1}^{2^{i-1}} (\pi/2 - \alpha_j)^2.
\end{equation}

Accordingly, the additional elastic energy required for bending a cluster of the order $i$ as a whole  is
\begin{widetext}
\begin{equation}
 \delta E_{\text{el}}^{(i)} = k\sum_{j=1}^{2^{i-1}} \left[ \left(\pi/2 -( \alpha_j + \delta\alpha_j )\right)^2 + \left( \pi/2 - (\alpha_j - \delta\alpha_j \right))^2 \right] - E_{\text{el}}^{(i)} = 2k\sum_{j=1}^{2^{i-1}} (\delta\alpha_j)^2 .
\end{equation}
\end{widetext}
The bending of a cluster as a whole leads to a displacement of the top of the bundle, characterized by the corresponding angular deviations $\delta\alpha_j$ of the participating plates. As an approximation we take $\delta\alpha_j=\delta\alpha$, which is valid if the lateral displacement of the top of the plates due to bending is small compared with the height $H$ of the plates.
This leads to
\begin{equation}
 \delta E_{\text{el}}^{(i)} = 2^ik(\delta\alpha)^2.
\end{equation}

\begin{figure}
  
 \parbox{0.15\columnwidth}{\includegraphics[width=0.15\columnwidth]{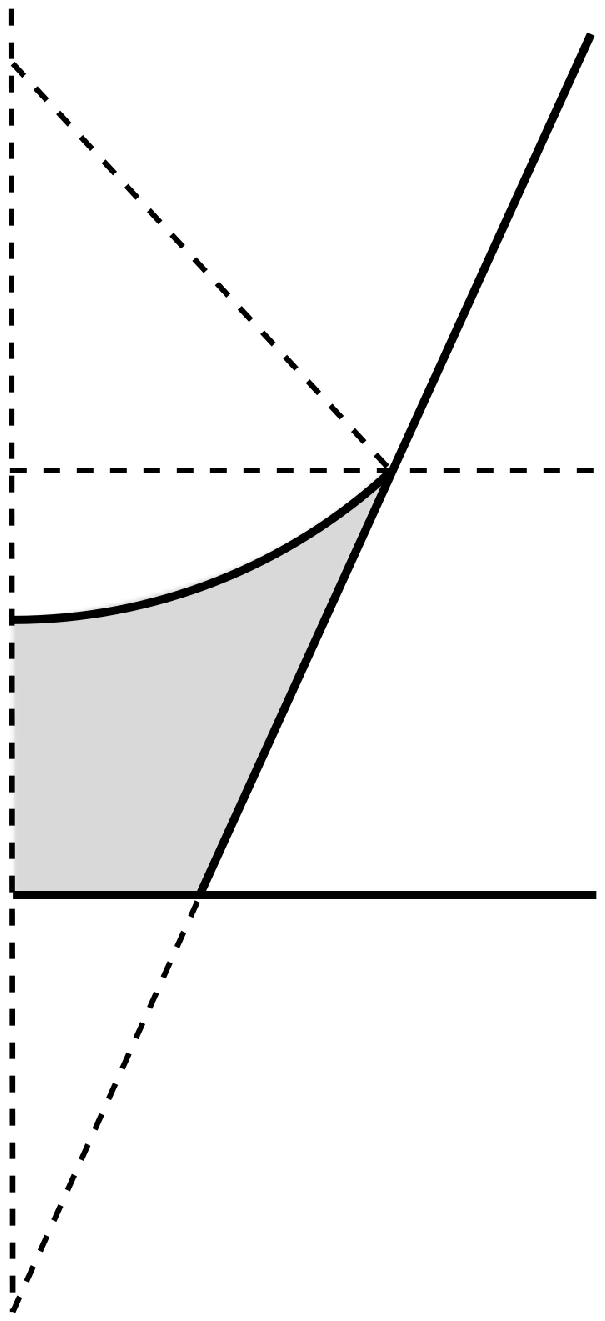}}
  =
  \parbox{0.15\columnwidth}{\includegraphics[width=0.15\columnwidth]{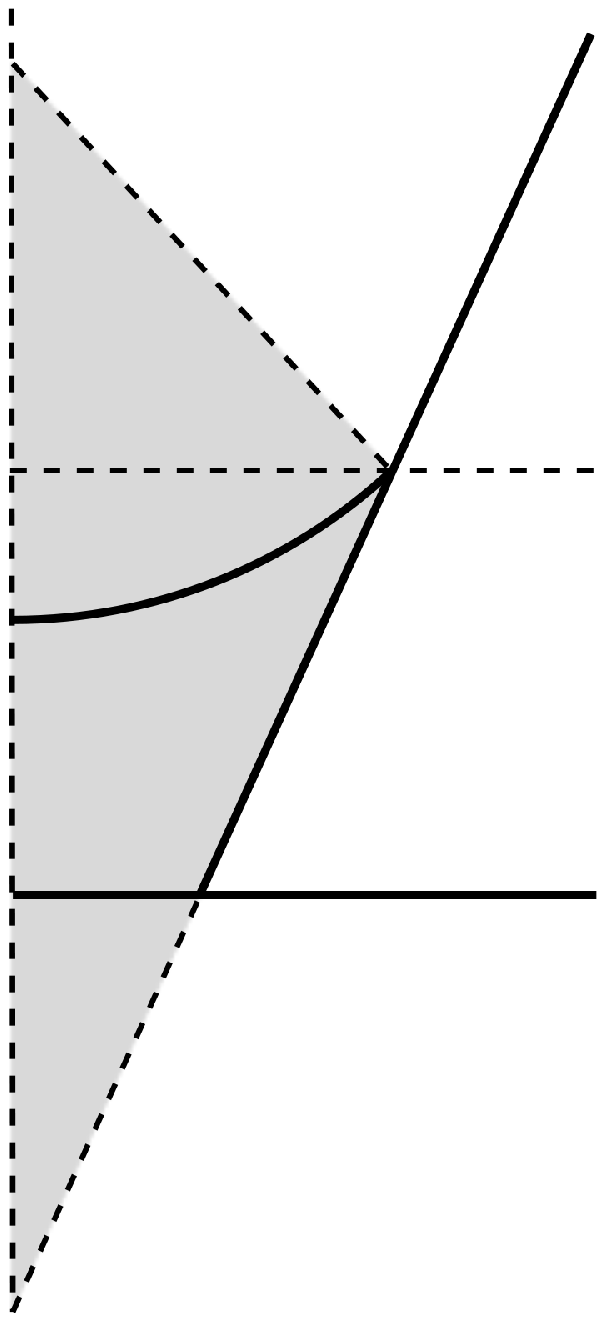}} --
  \parbox{0.15\columnwidth}{\includegraphics[width=0.15\columnwidth]{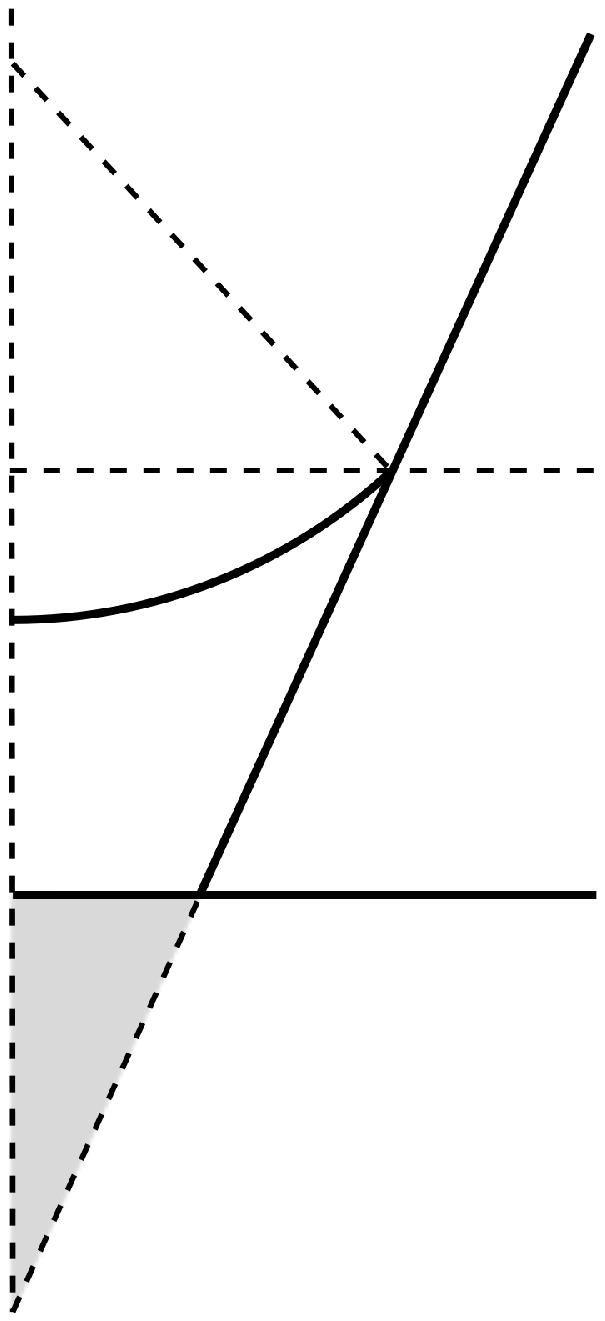}} --
  \parbox{0.15\columnwidth}{\includegraphics[width=0.15\columnwidth]{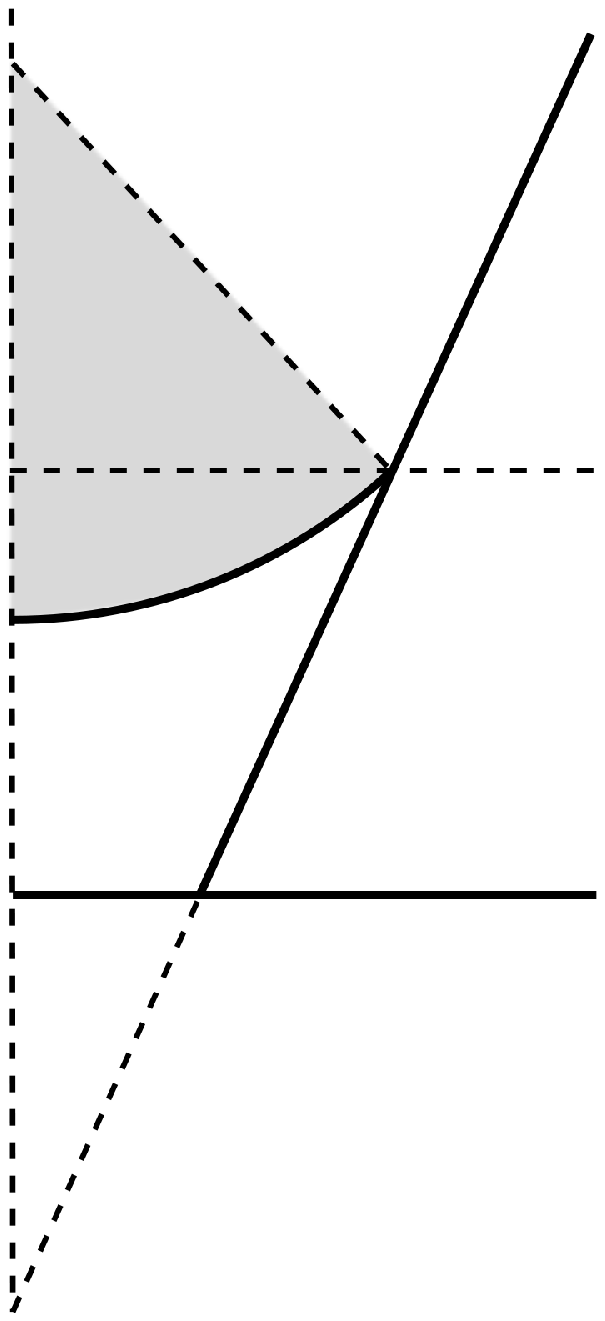}}

\caption{\label{volume}Representation of the volume of liquid adsorbed.}
\end{figure}

Concerning the contribution to the volume terms of the free energy one infers from 
Fig.~\ref{volume} that the volume of liquid (divided by the linear extension of the system in the translationally invariant direction) in a wedge after the $i^{\text{th}}$ clustering is
\begin{eqnarray}
 \Gamma_w^{(i)} & = & lh - h' L/2\nonumber\\
 & & - \left [ (\alpha_i-\theta) R^2
 - R^2\sin(\alpha_i-\theta)\cos(\alpha_i-\theta)\right]\\
  & = & R^2\left[ \frac{\sin(\alpha_i-\theta)\cos\theta}{\cos\alpha_i} - (\alpha_i-\theta) \right] -\frac{L^2}{4}\tan\alpha_i.\label{wedge}\nonumber
\end{eqnarray}
Likewise the surface contributions follow from inspection of Fig.~\ref{diagram}.
The surface area of the liquid-vapor interface is
\begin{equation}
 S_{\text{lv}}^{(i)} = 2(\alpha_i-\theta)R.
\end{equation}
The surface area of the liquid-solid interface is
\begin{equation}
  S_{\text{ls}}^{(i)} = L + \frac{2}{\cos\alpha_i}(R\sin(\alpha_i-\theta) - L/2).
\end{equation}
Finally, the surface area of the vapor-solid interface is
\begin{equation}
  S_{\text{vs}}^{(i)} = \frac{2H}{\sin\alpha_i} - S_{\text{ls}}^{(i)} + L.
\end{equation}

Adding these terms and using Laplace's equation (Eq.~(\ref{laplace})) and Young's equation
\begin{equation}
 \cos \theta = \frac{\sigma_{\text{vs}}-\sigma_{\text{ls}}}{\sigma}
\end{equation}
one obtains for the total free energy per number $N_{\text{c}}^{i}$ of clusters
\begin{widetext}
\begin{equation}
\label{last}
 E_i = \frac{\sigma^2}{\Delta \rho \Delta \mu}
    \left[
	(\alpha_i-\theta) - \frac{\cos\theta}{\cos\alpha_i}\sin(\alpha_i-\theta) - \frac{L^2}{4}\frac{(\Delta\rho\Delta\mu)^2}{\sigma^2}\tan\alpha_i + L\frac{\cos\theta}{\cos\alpha_i}\frac{\Delta\rho\Delta\mu}{\sigma}
     \right]  + E_{\text{el}}^{(i)}. \label{energycluster}
\end{equation}
\end{widetext}
In Eq.~\ref{last} we have omitted contributions which are independent of $\alpha_i$ and the contribution $\frac{2H}{\sin\alpha_i}$ to the vapor-solid surface energy. This latter term is an artifact generated by assuming a constant value of $H$ for the position of the top of the plates. Moreover we have dropped the contribution $\Gamma_{\text{c}}^{(i)}\Delta\rho\Delta\mu$ from the above expression. These terms might seem to be relevant as they depend on $\alpha_i$. However, the important dependence resides in $1/\cos\alpha_i$ because we always consider the values of $\alpha_i$ to be close to $\pi/2$ and thus the terms we have discarded are quantitatively unimportant. In the same spirit one could also drop the term proportional to $\alpha_i-\theta$, which will be canceled in further calculations anyway.

\end{document}